\documentclass[final,3p,times,sort&compress]{elsarticle}

\usepackage{epsfig}
\usepackage{amssymb}
\usepackage{amsthm}
\usepackage{color}
\usepackage{graphicx}
\usepackage[table]{xcolor}
\usepackage{lscape}
\usepackage{caption}
\usepackage{longtable}
\usepackage{soul}
\usepackage{amsmath,amssymb}
\usepackage{color}

\journal{Physica A}

\begin{document}

\begin{frontmatter}



\title{Rotating magnetoelectric effect in a ground state of a coupled spin-electron model on a doubly decorated square lattice}
\author[label1]{Hana \v Cen\v carikov\'a\corref{cor1}}
\ead{hcencar@saske.sk}
\cortext[cor1]{Corresponding author:}
\author[label2]{Jozef Stre\v{c}ka}

\address[label1]{Institute  of  Experimental  Physics,  Slovak   Academy   of Sciences, Watsonova 47, 040 01 Ko\v {s}ice, Slovakia}
\address[label2]{Department of Theoretical Physics and Astrophysics, Faculty of Science, P.~J. \v{S}af\' arik University, Park Angelinum 9, 040 01 Ko\v{s}ice, Slovakia}
\begin{abstract}
Exact analytical calculations are performed to study the rotating magnetoelectric effect in a ground state of a coupled spin-electron model on a doubly decorated square lattice with and without  presence of an external magnetic field. Novel spatially anisotropic magnetic ground states emergent due to a rotation in an external  electric field are found at three physically interesting  electron concentrations ranging from a quarter up to a half filling. In absence of the magnetic field existence of spatially anisotropic structures  requires a fractional electron concentration, where a significant influence of spatial orientation of an electric field  is observed.  It turns out that  the investigated model exhibits a rotating magnetoelectric effect at all three concentrations with one or two consecutive critical points in presence of magnetic field. At the same time, the rotating electric field has a significant effect on a critical value of an electrostatic potential, which can be enhanced or lowered upon changing  the electron hopping and the magnitude of an applied magnetic field. Finally, we have found an intriguing interchange of magnetic order between the horizontal and vertical directions driven by a rotation of the electric field, which is however destabilized upon strengthening of the magnetic field.
\end{abstract}

\begin{keyword}
Classical spin models, Rotating magnetoelectric effect, Exact analytical calculations, Phase transitions


\end{keyword}

\end{frontmatter}


\section{Introduction}
\label{s1}

The study of the magnetoelectric effect as well as materials with an electric-field controlled magnetism has been a subject of the longstanding interest in the condensed matter physics. At the beginning the  concerns of researchers in this field were attracted only sporadically, however, the discovery of novel functional materials~\cite{Fiebig,Nikitin,Tokura} has a dramatic enhancement of researcher interest in this specific area~\cite{Cao}. The main reason of such expansion relates to a wide range of practical applications, e.g., in  the spintronics, automation engineering, security, navigation or medicine~\cite{Prinz,Wolf,Son,Scott,Hur,Wu,Vopson,Sreenivasulu} and also  lies in  a necessity to deeper understand the processes responsible for the unconventional properties of these functional materials. 
Another motivation for enhanced research activities in this specific area is based on a requirement to find  novel  low-energy consuming and/or eco-friendly mechanisms preserving a functional character of used materials.  

In the present paper we would like to extend our previous study, where the novel concept about the rotating magnetoelectric effect has been introduced~\cite{Cenci2019}. Our considerations have been based on the analogy with a recent discovery of a rotating magnetocaloric effect, where the cooling or heating  is achieved through a spatial rotation of the  sample in a constant magnetic field instead of its moving in and out of a magnet or by actively changing the magnetic field~\cite{Nikitin}. As was shown during next few  years, the rotary magnetic refrigeration  observed in various anisotropic magnetic materials is technically more convenient and effective in comparison to its conventional counterpart~\cite{Zhang,Caro,Lorusso,Balli1,Balli3,Orendac,Moon}. Based on this idea we have applied an external electric field along the arbitrary spatial direction  lying in the plane of two-dimensional (2D) half-filled spin-electron system and we have observed that spatially modulated  influence of an electric field on charged particles  significantly influences the critical temperature of an investigated 2D system~\cite{Cenci2019}. The  detection of a substantial rotating magnetoelectric effect at a half filling motivated us to investigate an identical spin-electron model in other physically interesting electron concentrations to complement a rich spectrum of its unconventional properties~\cite{Cenci1,Cenci2,Cenci4}. Our further analyses will be exclusively limited to an investigation of a zero-temperature rotating magnetoelectric effect, which allows to examine it in the most authentic manner without the side effect of thermal fluctuations.

The outline of this paper is as follows. In Sec.~\ref{model} we will briefly introduce an investigated model together with a method used to study the rotating magnetoelectric effect.  The most interesting results demonstrating  existence of the aforementioned  phenomenon with and without the influence of external magnetic field will be detailed  discussed in Sec.~\ref{results}. Finally, the summary of the scientifically most significant  achievements will be mentioned in Sec.~\ref{conclusion}.

\section{Model and Method}
\label{model}
\begin{figure}[t]
\begin{center}
$(a)$\hspace*{4cm}$(b)$\hspace*{6cm}\\
{\includegraphics[width=0.2\textwidth,height=0.15\textheight,trim=2cm 5cm 9cm 12cm, clip]{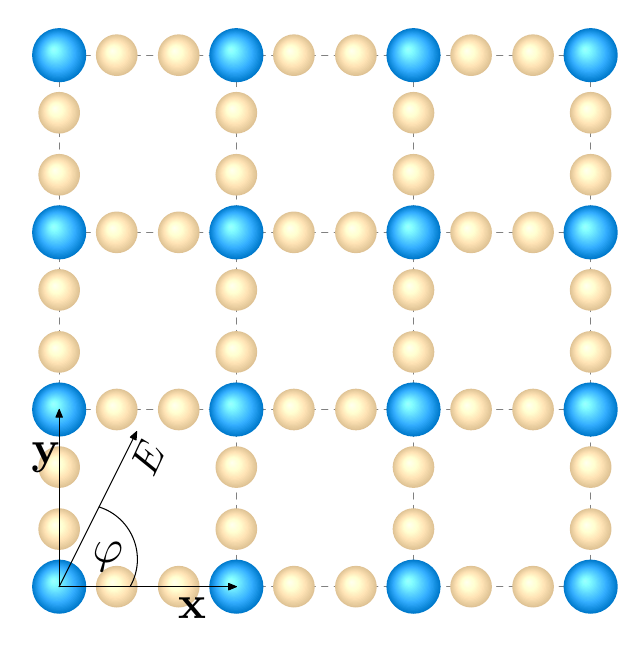}}\hspace*{0.8cm}
{\includegraphics[width=0.2\textwidth,height=0.15\textheight,trim=2.8cm 6.2cm 12.5cm 14.5cm, clip]{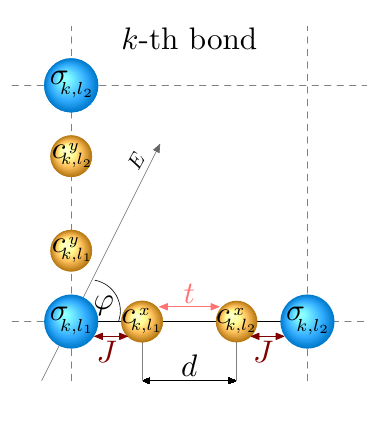}}
\caption{\small A schematic representation of a coupled spin-electron system on a 2D doubly decorated square lattice  $(a)$ supplemented with a specific detail of the $k$-th unit cell $(b)$. Large (blue) balls represent the location of Ising spins and small (yellow) balls denote the location of decorating sites over which the mobile electrons are delocalized. All relevant interactions entering into the model Hamiltonian are also visualized.}
\label{fig1}
\end{center}
\end{figure}

The investigated coupled spin-electron model on a doubly decorated square lattice consists of localized Ising spins $\hat{\sigma}^z$  with two possible eigenvalues ${\sigma}^z\!=\!\pm 1$ and the mobile electrons delocalized over all bonds lying in  between two nearest-neighbor Ising spins as displayed in Fig.~\ref{fig1}. 
The local character of all present interactions allows us to decompose the total Hamiltonian $\hat{\cal{H}}$ into the $N$ horizontal  ($\hat{\cal{H}}^x_k$) and $N$  vertical ($\hat{\cal{H}}^y_k$) bond Hamiltonians, which can be defined through the following formula
\allowdisplaybreaks
\begin{align}
\hat{\cal H}^{\delta}_k=&-t(\hat{c}^{\dagger,\delta}_{k,l_1,\uparrow}\hat{c}^{\delta}_{k,l_2,\uparrow}\!+\!\hat{c}^{\dagger,\delta}_{k,l_1,\downarrow}\hat{c}^{\delta}_{k,l_2,\downarrow}\!+\!h.c.)
-
\sum_{\alpha=1,2}\left(J\hat\sigma^{z,\delta}_{k,l_{\alpha}}\!+\!h\right)\left(\hat{n}^{\delta}_{k,l_{\alpha},\uparrow}\!-\!\hat{n}^{\delta}_{k,l_{\alpha},\downarrow}\right)
\nonumber
\\
&-\sum_{\alpha=1,2}\left[\frac{h}{4}\hat\sigma^{z,\delta}_{k,l_{\alpha}}
\!-\!(-1)^{\alpha}V_{\delta}\sum_{\gamma=\uparrow,\downarrow}\hat{n}^{\delta}_{k,l_{\alpha},\gamma}\right]\!-\!\mu\hat{n}^{\delta}_{k}\;,
\label{eq1}
\end{align}
where $\delta\!=\!x,y$ and $\hat{c}^{\dagger,\delta}_{k,l_{\alpha},\gamma}$, $\hat{c}^{\delta}_{k,l_{\alpha},\gamma}$ ($\alpha\!=\!1,2$; $\gamma\!=\!\uparrow,\downarrow$) denote the creation and annihilation fermionic operators of the mobile electrons from the $k$-th bond  with the respective number operators $\hat{n}^{\delta}_{k,l_{\alpha},\gamma}\!=\!\hat{c}^{\dagger,\delta}_{k,l_{\alpha},\gamma}\hat{c}^{\delta}_{k,l_{\alpha},\gamma}$  determining occupation of a decorating site $l_{\alpha}$ through a mobile electron with a spin $\gamma$. The sum of all partial number operators $\hat{n}^{\delta}_{k,l_{\alpha},\gamma}$  commonly defines the total number operator at $k$-th bond, $\hat{n}^{\delta}_k\!=\!\sum_{\alpha=1,2}(\hat{n}^{\delta}_{k,l\alpha,\uparrow}\!+\!\hat{n}^{\delta}_{k,l\alpha,\downarrow})$. The relevant interactions entering into the horizontal as well as vertical bond Hamiltonians~(\ref{eq1}) correspond to the kinetic energy of mobile electrons  modulated by the hopping amplitude $t$, the Ising-type exchange interaction $J$ between the nearest-neighbor  localized Ising spins and mobile electrons. Parameters $h$ and $V_{\delta}$ take into account an influence of the external magnetic and electric fields, respectively. While the magnetic field  affects simultaneously electron as well as spin subsystem, the electric field $E$  acts exclusively on the subsystem of mobile electrons with a charge $|e|$. Imposing the distance $d$ between nearest-neighbor decorating sites allows us to unambiguously determine the magnitude of the electrostatic potential $V$ ($V\!=\!E|e|d/2$). The difference between  the horizontal and vertical bond Hamiltonians~(\ref{eq1}) originates from a different contribution of the electrostatic potential $V$ on the horizontal ($V_{x}$) and vertical ($V_y$) bonds, which can be easily tuned upon varying an inclination of the electric field from the global frame axis $x$, as determined by the polar angle $\varphi$ (see Fig.~\ref{fig1}). Owing to this fact, the horizontal and vertical contribution of the electrostatic energy can be expressed as  $V_x\!=\!V\cos\varphi$ and $V_y\!=\!V\sin\varphi$, respectively. The last term $\mu$ entering into the Eq.~(\ref{eq1}) is the standard chemical potential, which controls the number of mobile electrons per bond.

In order to analyze the rotating magnetoelectric effect at zero temperature we have at first determined the magnetic phase diagrams as a function of the polar angle $\varphi$. The energy of each magnetic configuration entering into the ground-state phase diagram is defined as a sum of the lowest-energy eigenstates over all $N$ pairs of bonds in the horizontal (${\varepsilon}^{x}_{k_i}, i\!=\!1,2,\dots,16$) and vertical (${\varepsilon}^{y}_{k_i}, i\!=\!1,2,\dots,16$)  direction,  ${\varepsilon}\!=\!\sum_{k=1}^{N}\left({\varepsilon}^{x}_k\!+\!{\varepsilon}^y_k\right)$. It should be mentioned that the electron occupation $\langle \hat{n}_k^{\delta}\rangle$ can be different for  two orthogonal spatial directions. Since the different occupation number $\langle \hat{n}_k^{\delta}\rangle$ is allowed, it is convenient to define the mean electron concentration $\rho$ through the relation $\rho\!=\!(\langle \hat{n}_k^{x}\rangle\!+\!\langle \hat{n}_k^{y}\rangle)/2$. The lowest-energy  eigenstates in both spatial directions can be  derived from the  sixteen energy eigenvalues ${\varepsilon}_{k_i}^{\delta}$ ($i\!=\!1,2,\dots,16$) obtained by an exact diagonalization procedure  (reported in detail in  Ref.~\cite{Cenci0}) for all available configurations of Ising spins $\sigma^{\delta}_{k,l_1}$ and $\sigma^{\delta}_{k,l_2}$ 
\allowdisplaybreaks
\begin{align}
\langle \hat{n}_k^{\delta}\rangle\!=\!0&:{\varepsilon}^{\delta}_{k_1}\!=\!-\frac{h}{2}A_{+}^{\delta},\nonumber\\
\langle \hat{n}_k^{\delta}\rangle\!=\!1&: {\varepsilon}^{\delta}_{k_2,\,k_3}\!=\!-(JA_{+}^{\delta}\!+\!h)\!-\!\frac{h}{2}A_{+}^{\delta}\!\mp\! B^{\delta}_{+}\!-\!\mu,\nonumber\\
&:{\varepsilon}^{\delta}_{k_4,\,k_5}\!=\!(JA_{+}^{\delta}\!+\!h)\!-\!\frac{h}{2}A_{+}^{\delta}\!\mp\! B^{\delta}_{-}\!-\!\mu,\nonumber\\
\langle \hat{n}_k^{\delta}\rangle\!=\!2&: {\varepsilon}^{\delta}_{k_6,\,k_7}\!=\!\mp2(JA_{+}^{\delta}\!+\!h)\!-\!\frac{h}{2}A_{+}^{\delta}\!-\!2\mu,\nonumber\\
&:{\varepsilon}^{\delta}_{k_8,\,k_9}\!=\!-C^{\delta}_{\pm}\!-\!\frac{h}{2}A_{+}^{\delta}\!-\!2\mu,\label{eq2}\\
&:{\varepsilon}^{\delta}_{k_{10},\,k_{11}}\!=\!C^{\delta}_{\pm}\!-\!\frac{h}{2}A_{+}^{\delta}\!-\!2\mu,\nonumber\\
\langle \hat{n}_k^{\delta}\rangle\!=\!3&:{\varepsilon}^{\delta}_{k_{12},\,k_{13}}\!=\! -(JA_{+}^{\delta}\!+\!h)\!-\!\frac{h}{2}A_{+}^{\delta}\!\pm\! B^{\delta}_{-}\!-\!3\mu,\nonumber\\
&: {\varepsilon}^{\delta}_{k_{14},\,k_{15}}\!=\! (JA_{+}^{\delta}\!+\!h)\!-\!\frac{h}{2}A_{+}^{\delta}\!\pm\!B^{\delta}_{+}\!-\!3\mu,\nonumber\\
\langle \hat{n}_k^{\delta}\rangle\!=\!4&:{\varepsilon}^{\delta}_{k_{16}}\!=\!-\frac{h}{2}A_{+}^{\delta}\!-\!4\mu.
\nonumber
\end{align}
Here,  $A_{\pm}^{\delta}\!=\!(\sigma^{\delta}_{k,l_1}\!\pm\!\sigma^{\delta}_{k,l_2})/2$, $B^{\delta}_{\pm}\!=\!\sqrt{\left[JA_{-}^{\delta}\!\pm\! V_{\delta}\right]^2\!+\!t^2}$ and $C^{\delta}_{\pm}\!=\!\sqrt{2}\sqrt{B^{\delta}_{+}(B^{\delta}_{+}\!\pm\!B^{\delta}_{-})\!-\!2JA_{-}^{\delta}V_{\delta}}$. The lowest-energy eigenstate is then unambiguously determined by the corresponding eigenvector  $|\psi\rangle\!=\!\prod_{k=1}^{N} |\psi\rangle_k^x|\psi\rangle_k^y$, which minimizes the overall energy. Of course, the total normalized magnetization $m_{tot}$ of the investigated spin-electron model (\ref{eq1}) is given by 
\begin{align}
m_{tot}&=\frac{1}{2}\sum_{\delta=x,y}m_{tot}^{\delta}\!=\!\frac{1}{2}\sum_{\delta=x,y}\sum_{k=1}^N\frac{\left(m_{k,I}^{\delta}\!+\!m_{k,e}^{\delta}\right)}{\left(1\!+\!\langle \hat{n}_k^{\delta}\rangle\right)}\;.
\label{eq3}
\end{align}
 For a completeness, the sublattice   magnetizations $m^{\delta}_{k,I}$ and  $m^{\delta}_{k,e}$ of the localized Ising spins and mobile electrons emergent  in Eq.~(\ref{eq3}) are defined as follows
\begin{align}
m^{\delta}_{k,I}&=\frac{\langle\hat{\sigma}^{\delta}_{k,l_1}\rangle\!+\!\langle\hat{\sigma}^{\delta}_{k,l_2}\rangle}{2}\;,
\nonumber\\
m^{\delta}_{k,e}&=\sum_{\alpha=1,2}\left(\langle \hat{n}^{\delta}_{k,l_{\alpha},\uparrow}\rangle-\langle\hat{n}^{\delta}_{k,l_{\alpha},\downarrow}\rangle\right)\;.
\label{eq4}
\end{align}
where $\delta=x,y$.
\section{Results and discussion}
\label{results}
Before discussing the most interesting results  it should be pointed out that all further analyzes will be performed for a ferromagnetic Ising interaction $J\!>\!0$, because the transformation $J\!\to\!-J$ results in a trivial interchange of a relative spin orientation of the mobile electrons with respect to their nearest Ising spins. To reduce a  parametric space, we have applied a restriction for a polar angle $\varphi$  from $\varphi\!=\!0$ to $\varphi\!=\!\pi/4$ in all subsequent analyzes, since the higher values of $\varphi$ only result in symmetry-related magnetic structures obtained by a rotation in $xy$-plane by the angle $\pi/2$. The restriction to 
$\langle\hat{n}_k^{\delta}\rangle\!\leq\!2$ is further applied when taking into account a validity of particle-hole symmetry.

First, let us take a closer look at the particular case without the external magnetic field $h/J$. Besides three spatially isotropic ground states (A, C and F) already reported in our previous work~\cite{Cenci2015}, the non-zero electric field may stabilize other four spatially anisotropic ones (B, D, E and E$^*$), which can be defined through a specific combination of three eigenvectors
\begin{align}
|\textrm{0}\rangle_k^{\delta}&=|1\rangle_{\sigma_{k,l_1}}\!\otimes\!|0,0\rangle_k\!\otimes\!|1\rangle_{\sigma_{k,l_2}}\;,
\nonumber\\
|\textrm{I}\rangle_k^{\delta}&=|1\rangle_{\sigma_{k,l_1}}\!\otimes\!\left(\cos a_1^{\delta}|\uparrow,0\rangle_k\!+\!\sin a_1^{\delta}|0,\uparrow\rangle_k\right)\!\otimes\!|1\rangle_{\sigma_{k,l_2}}\;,
\nonumber\\ 
&\tan a_1^{\delta}=\left(\sqrt{{V_{\delta}}^2\!+\!t^2}\!-\!V_{\delta}\right)/t\;,
\nonumber\\
|\textrm{II}\rangle_k^{\delta}&=|1\rangle_{\sigma_{k,l_1}}\!\otimes\!\left(a_2^{\delta}|\uparrow,\downarrow\rangle_k\!+\!b_2^{\delta}|\downarrow,\uparrow\rangle_k\!+\!c_2^{\delta}|\uparrow\downarrow,0\rangle_k\right.
\nonumber\\
&+\left.d_2^{\delta}|0,\uparrow\downarrow\rangle_k\right)\!\otimes\!|-1\rangle_{\sigma_{k,l_2}}\;,
\nonumber\\
a_2^{\delta}&=2t\xi_{\delta}(\xi_{\delta}\!+\!2J)/\eta_{\delta}\;,
\nonumber\\
b_2^{\delta}&=-a^{\delta}_2(\xi_{\delta}\!-\!2J)/(\xi_{\delta}\!+\!2J)\;,
\label{eq5}\\
c_2^{\delta}&=a^{\delta}_2(\xi_{\delta}\!-\!2J)(\xi_{\delta}\!+\!2V_{\delta})/2t\xi_{\delta}\;,
\nonumber\\
d_2^{\delta}&=c^{\delta}_2(\xi_{\delta}\!-\!2V_{\delta})
/(\xi_{\delta}\!+\!2V_{\delta})\;,
\nonumber\\
\xi_{\delta}&=\sqrt{2}\left[(J^2\!+\!V_{\delta}^2\!+\!t^2)\!+\! \left(\frac{(J\!+\!V_{\delta})^2\!+\!t^2}{\left((J\!-\!V_{\delta})^2\!+\!t^2\right)^{-1}}\right)^{1/2}\right]^{1/2}\;,
\nonumber\\
\eta_{\delta}&=\sqrt{2}\left[4t^2{\xi_{\delta}}^2({\xi_{\delta}}^2\!+\!4J^2)
\!+\!\frac{({\xi_{\delta}}^2\!-\!4J^2)^2}{({\xi_{\delta}}^2\!+\!4V_{\delta}^2)^{-1}}\right]^{1/2}\;,
\nonumber\\
\delta&=x,y\;.\nonumber
\end{align}
A complete list of all possible ground states is reported in  Tab.~\ref{tab1}, where individual ground states sorted according to the  electron concentration $\rho$ are supplemented with the definition of their notation, ground-state energy (${\varepsilon}_k$), the explicit form of the relevant eigenvector ($|\psi\rangle_k$) and the total magnetization ($m_{tot}$).
\begin{table}[b!]
\begin{center}
\begin{tabular}{l|c|c|l|c}
\hline
\rowcolor[gray]{0.7}
$\rho$&{Phase} &${\varepsilon}_k$&$|\psi\rangle_k$&$m_{tot}$\\
\hline\hline
0 & A & $-h$&$|{\rm 0}\rangle_k^x\,|{\rm 0}\rangle_k^y$&1 \\
\hline
0.5&B &$-J-2h-\mu-\omega_x$& $|{\rm I}\rangle_k^x\,\,|{\rm 0}\rangle_k^y$&1\\
\hline
1&C &$-2J-3h-2\mu-\omega_x-\omega_y$& $|{\rm I}\rangle_k^x\,\,|{\rm I}\rangle_k^y$&1\\
1&D&$-h/2-2\mu-\Omega_x$ & $|{\rm II}\rangle_k^x|{\rm 0}\rangle_k^y$&1/2\\
\hline
1.5&E &$-J-3h/2-3\mu-\omega_x-\Omega_y$ & $|{\rm I}\rangle_k^x\,\,|{\rm II}\rangle_k^y$&1/2\\
1.5&E$^*$ &$-J-3h/2-3\mu-\Omega_x-\omega_y$ & $|{\rm II}\rangle_k^x|{\rm I}\rangle_k^y$&1/2\\
\hline
2& F &$-4\mu-\Omega_x -\Omega_y$ & $|{\rm II}\rangle_k^x|{\rm II}\rangle_k^y$&0\\
\hline
\end{tabular}
\caption{\normalsize A complete list of all  ground states of a coupled spin-electron  model on a doubly decorated square lattice for the zero-magnetic-field case. Here ${\Omega_{x(y)}}\!=\!\sqrt{2(J^2\!+\!\omega_{x(y)}^2)\!+\!2\sqrt{(J^2\!+\!\omega_{x(y)}^2)^2\!-\!4J^2V_{x(y)}^2}}$ and ${\omega_{x(y)}}\!=\!\sqrt{V_{x(y)}^2\!+\!t^2}$\;.}
\label{tab1}
\end{center}
\end{table}
As one can see, 
\begin{figure*}[t!]
\begin{center}
{\includegraphics[width=0.45\textwidth,height=0.25\textheight,trim=3.1cm 9.3cm 3.1cm 9.3cm, clip]{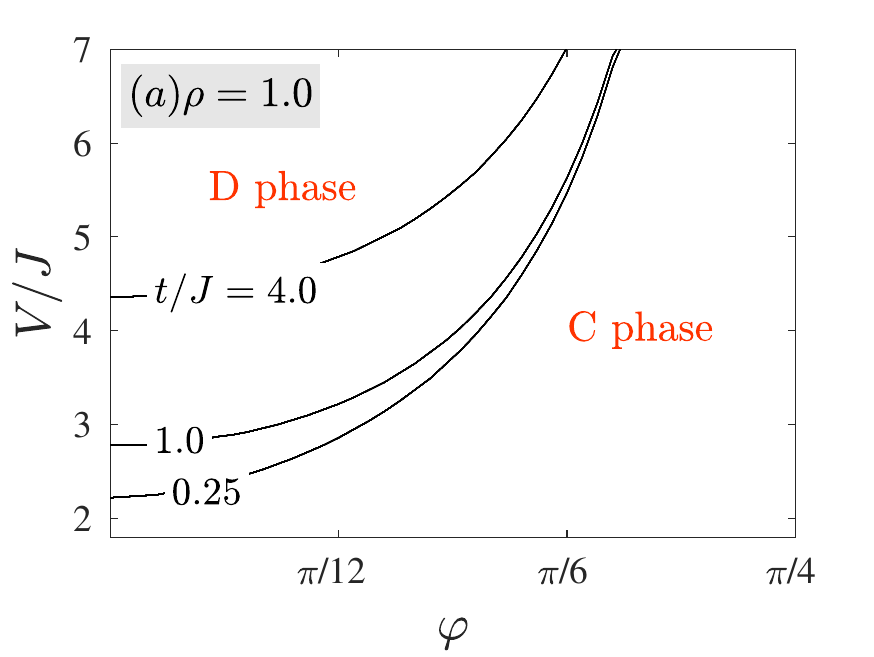}}
\hspace*{1cm}
{\includegraphics[width=0.45\textwidth,height=0.25\textheight,trim=3.1cm 9.3cm 3.1cm 9.3cm, clip]{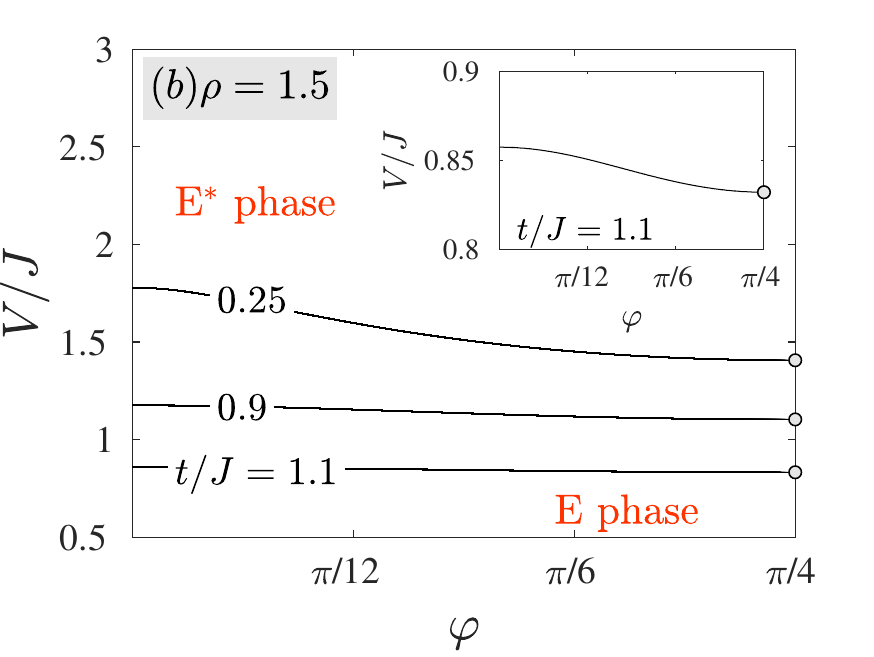}}
\caption{\small Ground-state phase diagram  in the $\varphi-V/J$ plane between two spatially different ground states with the total electron occupation $\rho\!=\!1$ $(a)$ and  $\rho\!=\!1.5$ $(b)$ for selected values of the  electron hopping $t/J$ in absence of an external magnetic field $h/J$. Light gray circles in the panel $(b)$ denote the fact that the phases E and E$^*$ coexist together for arbitrary $V/J$ if $\varphi\!=\!\pi/4$.} 
\label{fig2}
\end{center}
\end{figure*}
 two spatially inhomogeneous ground states may  exist due to a spatial    modulation of the electric field for two  electron concentrations  $\rho\!=\!1$ and $\rho\!=\!1.5$. The ground-state phase diagrams in the angle versus a relative strength of the electric field ($\varphi-V/J$) plane are presented in Fig.~\ref{fig2}.  It is evident that phase boundaries  between two different ground states exhibit a monotonic but non-constant angular dependence unambiguously demonstrating presence of the rotating magnetoelectric effect.  It also follows from Fig.~\ref{fig2} that  the rotating magnetoelectric effect  significantly depends on a relative  strength of the  hopping amplitude $t/J$. While  the strengthened  hopping amplitude $t/J$  shifts the relevant phase boundary between the ground states C and D to  higher values of the electric field $V/J$ at  the quarter electron filling ($\rho\!=\!1$), the opposite trend is observed for the phase boundary between the ground states E and E$^*$ at the  electron concentration $\rho\!=\!1.5$.  It is interesting to note that  the increasing  hopping term $t/J$ reduces for the electron concentration $\rho\!=\!1.5$ the rotating magnetoelectric effect as a consequence of a suppression of the spatially anisotropic phase E. Above the critical value of the electron hopping, numerically identified as $t_c/J\!\approx\!1.305$,  the  phase E is completely replaced by the phase E$^*$ and  the rotating magnetoelectric effect completely  vanishes. 
Moreover,  the angular change  of a spatial orientation of the external electric field ($\varphi\!<\!\pi/4$) allows to discontinuously swap   magnetic orders of a coupled spin-electron system for the electron concentration $\rho\!=\!1.5$ and relatively weak electron hopping $t/J\!\lessapprox\!1.305$ on the horizontal and vertical directions (E$\leftrightarrows$E$^*$), in spite of the fact that  the  total magnetization remains unchanged. 
\begin{figure}[t!]
\begin{center}
{\includegraphics[width=0.45\textwidth,trim=3.1cm 9.3cm 3.1cm 9.3cm, clip]{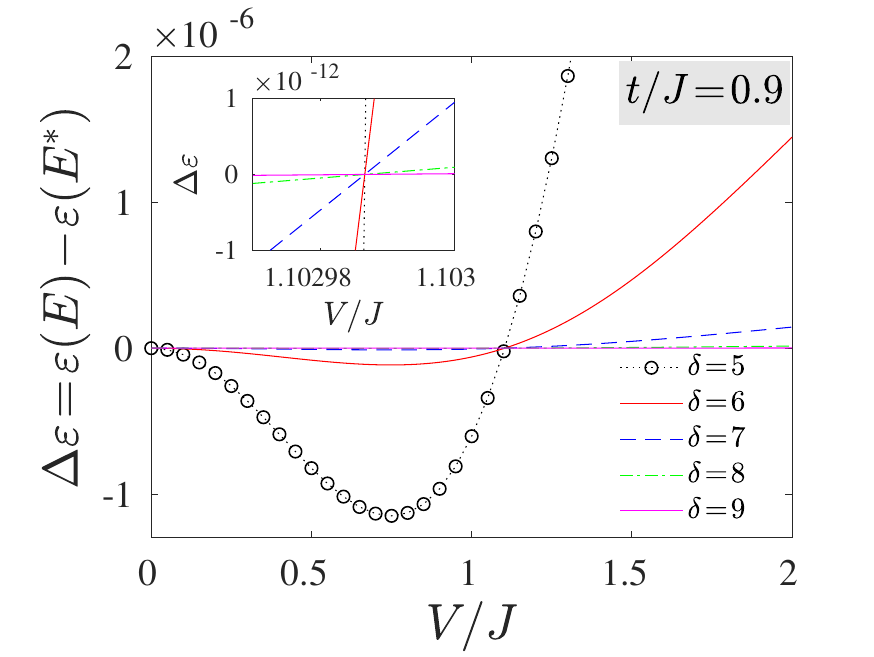}}
\caption{\small Energy difference $\Delta {\varepsilon}\!=\!{\varepsilon(E)}\!-\!{\varepsilon(E^*)}$  as a function of the electric field $V/J$, calculated for $t/J\!=\!0.9$ and $h/J\!=\!0$ for five different angular variables $\varphi\!=\!\pi/4\!-\!1\!\times\!10^{-\delta}$. Inset shows the details of  behaviour in the proximity of the transition electric field $V/J$.} 
\label{fig3}
\end{center}
\end{figure}
 For a completeness we note that for the specific orientation of the external electric field $\varphi\!=\!\pi/4$ both phases E and E$^*$ become energetically equivalent, because the spatial anisotropy vanishes and the energies of the single and doubly occupied bonds are identical for arbitrary  strength of the electric field $V/J$ (visualized as the light gray circles in Fig.~\ref{fig2}$(b)$). However, Fig.~\ref{fig3} demonstrates that arbitrary small but non-zero deviation from the most symmetric case $\varphi\!=\!\pi/4$   induces the phase transition E-E$^*$ at relatively high value of the electric field $V/J$, on assumption that $t/J\!\lessapprox\!t_c/J$.
\\

\begin{figure}[t!]
\begin{center}
{\includegraphics[width=0.45\textwidth,height=0.25\textheight,trim=3.1cm 9.3cm 3.1cm 9.3cm, clip]{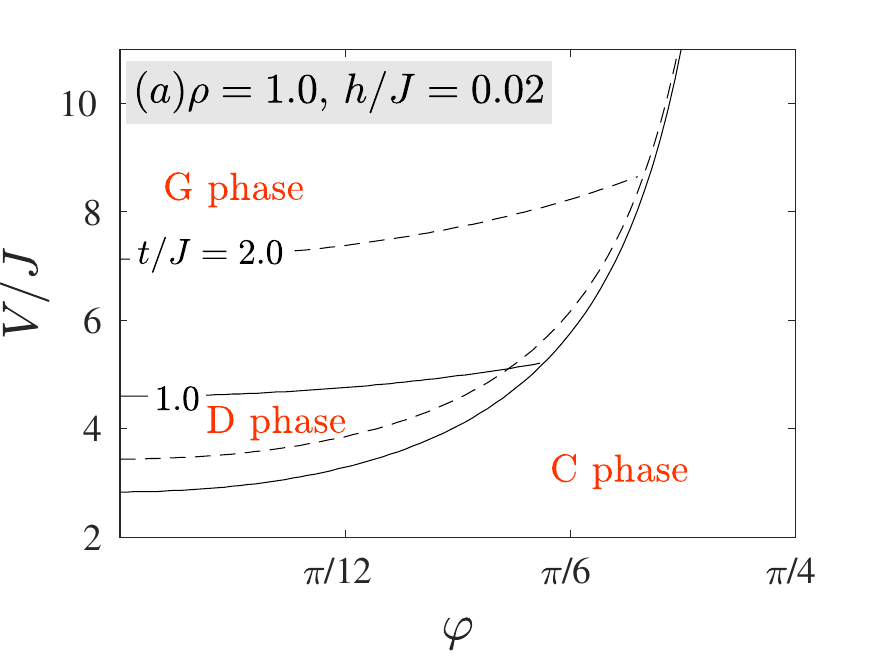}}
{\includegraphics[width=0.45\textwidth,height=0.25\textheight,trim=3.1cm 9.3cm 3.1cm 9.3cm, clip]{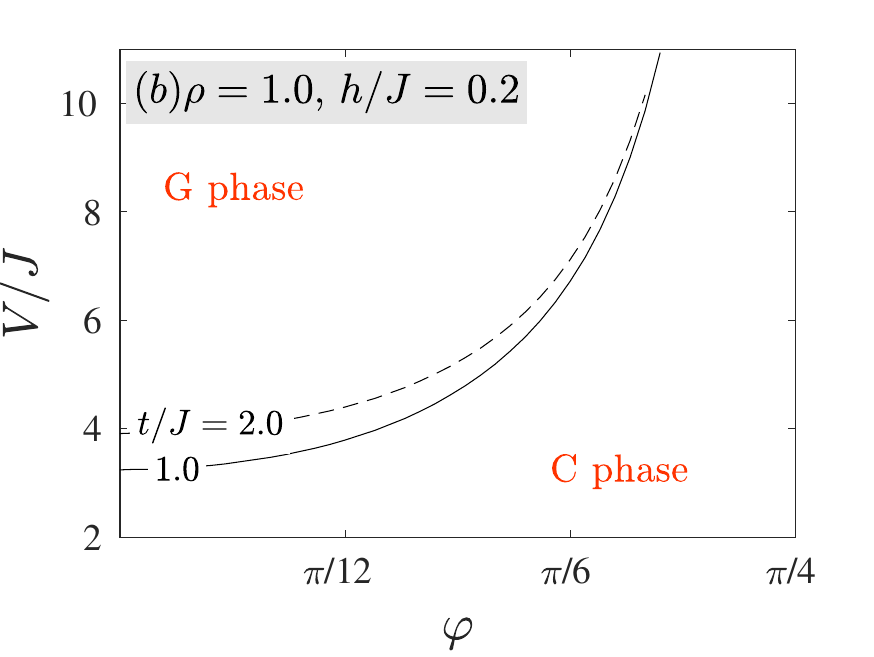}}
\caption{\small Ground-state phase diagram  in the $\varphi-V/J$ plane at a quarter filling $\rho\!=\!1.0$ for two different  values of the external magnetic field $h/J\!=\!0.02$ $(a)$ and $0.2$ $(b)$ and two different values of the  hopping amplitude $t/J\!=\!1.0$ (solid lines) and $t/J\!=\!2.0$ (dashed lines).} 
\label{fig4}
\end{center}
\end{figure}

\begin{figure*}[t!]
\begin{center}
{\includegraphics[width=0.45\textwidth,trim=3.1cm 9.3cm 3.1cm 9.3cm, clip]{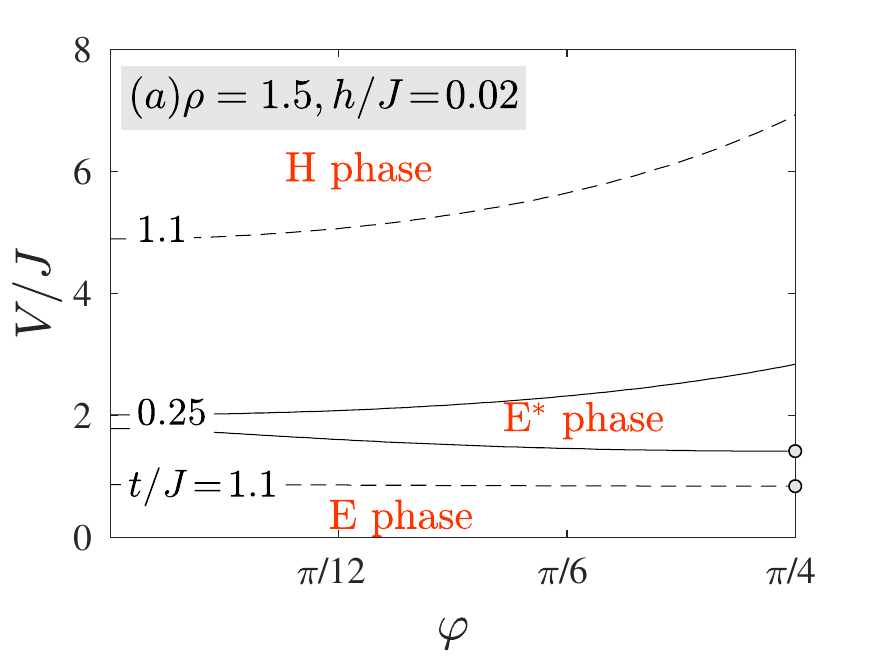}}
\hspace*{1cm}
{\includegraphics[width=0.45\textwidth,trim=3.1cm 9.3cm 3.1cm 9.3cm, clip]{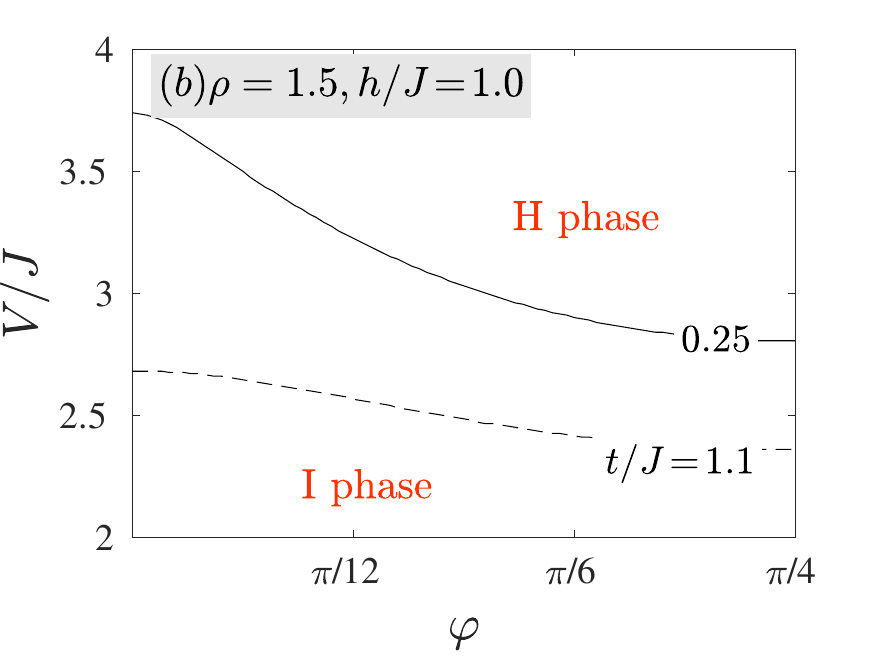}}
{\includegraphics[width=0.45\textwidth,trim=3.1cm 9.3cm 3.1cm 9.3cm, clip]{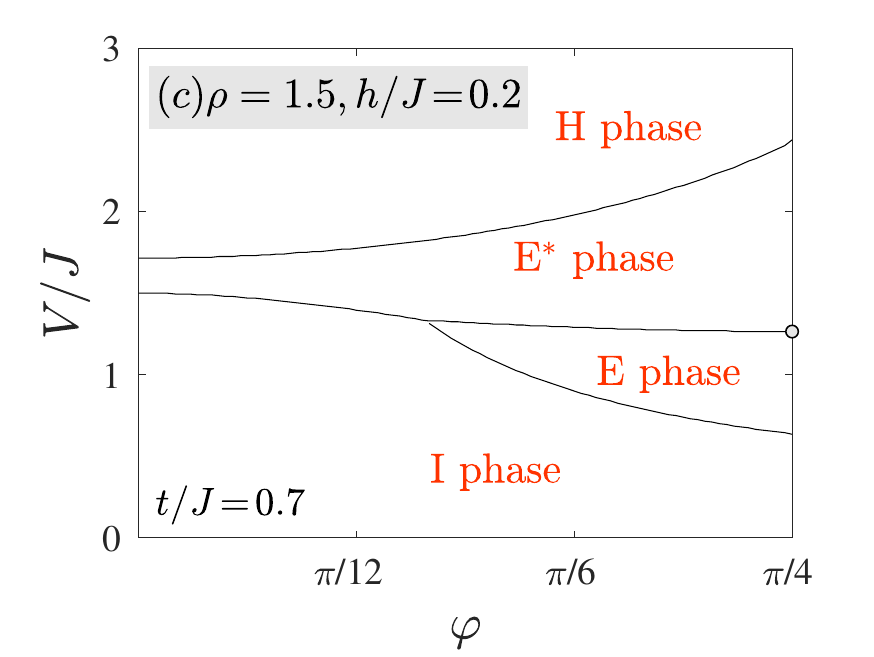}}
\hspace*{1cm}
{\includegraphics[width=0.45\textwidth,trim=3.1cm 9.3cm 3.1cm 9.3cm, clip]{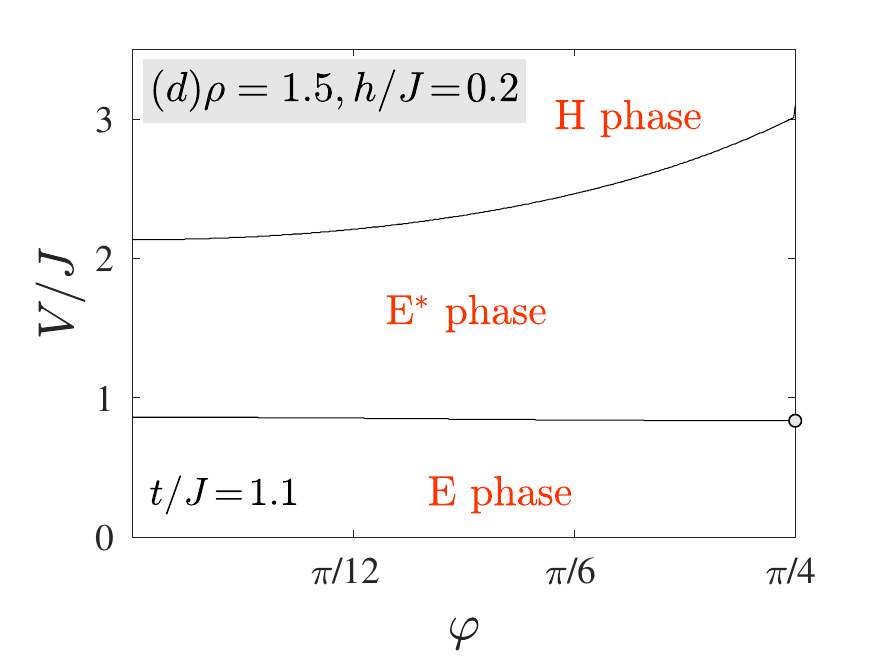}}
\caption{\small 
Ground-state phase diagram  in the $\varphi-V/J$ plane at a fractional electron  filling $\rho\!=\!1.5$ for three different  values of the external magnetic field $h/J\!=\!0.02$ $(a)$, 1.0  $(b)$  and $0.2$ $(c)$-$(d)$. Different  values of the  hopping amplitude $t/J$ are taken into account.  Light gray circles in panels $(a), (c)$ and $(d)$ denote the fact that the phases E and E$^*$ coexist together for arbitrary $V/J$ if $\varphi\!=\!\pi/4$.
} 
\label{fig5}
\end{center}
\end{figure*}

\begin{figure}[t!]
\begin{center}
{\includegraphics[width=0.33\textwidth,trim=3cm 9cm 3cm 9.cm, clip]{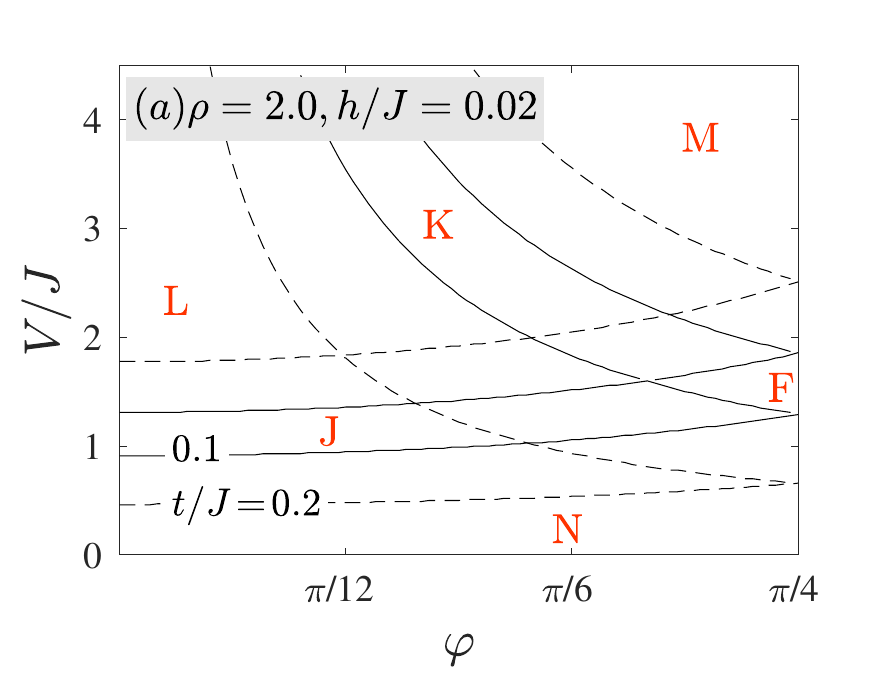}}
{\includegraphics[width=0.33\textwidth,trim=3cm 9cm 3cm 9.cm, clip]{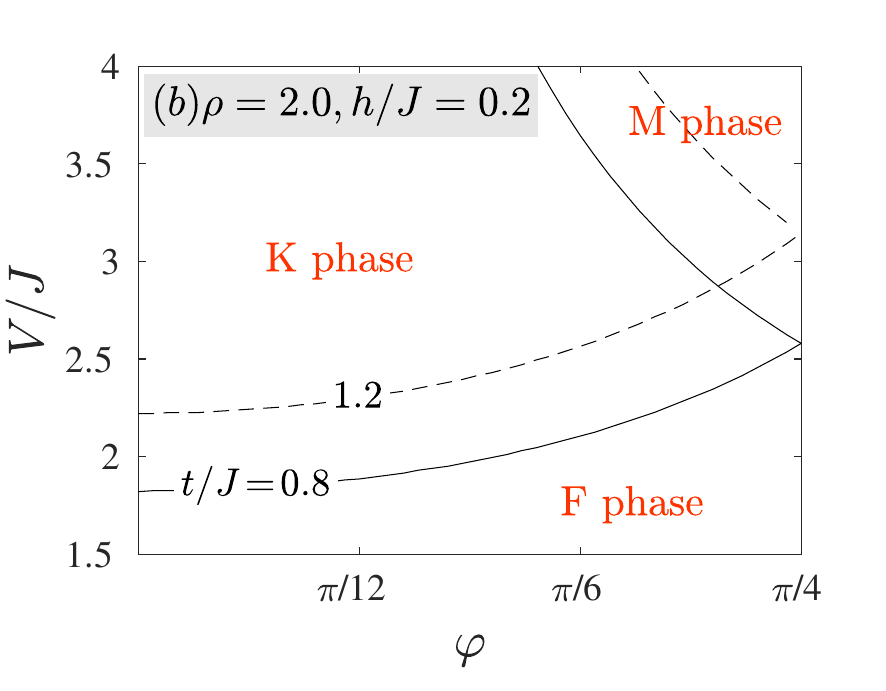}}
{\includegraphics[width=0.33\textwidth,trim=3cm 9cm 3cm 9.cm, clip]{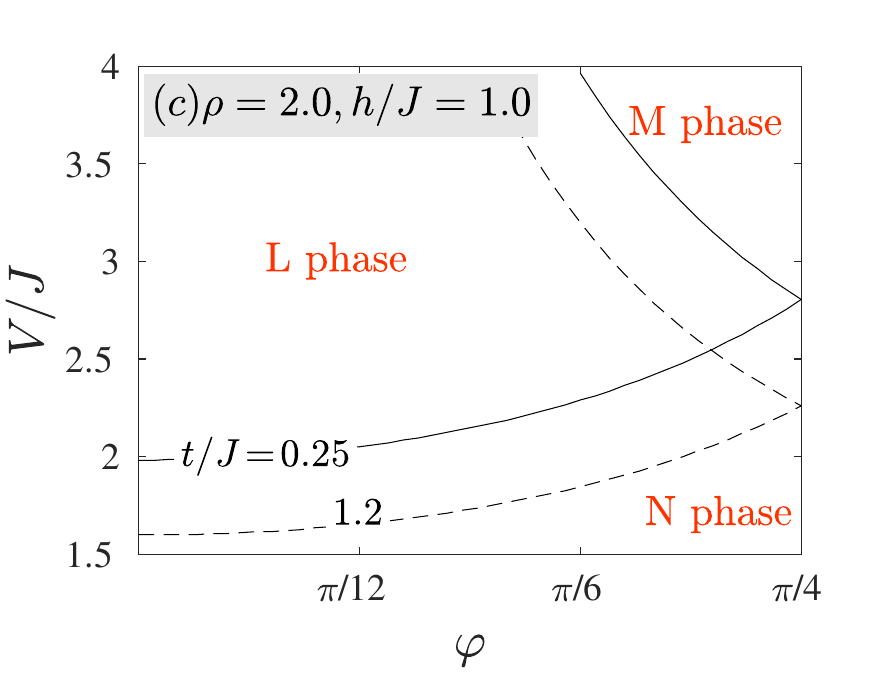}}
\caption{\small Ground-state phase diagram  in the $\varphi-V/J$ plane at a half   filling $\rho\!=\!2.0$ for three different  values of the external magnetic field $h/J\!=\!0.02$ $(a)$, 0.2  $(b)$  and 1.0 $(c)$. Different  values of the  hopping amplitude $t/J$ are taken into account.
} 
\label{fig6}
\end{center}
\end{figure}

As a next step, let us examine  existence of the rotating magnetoelectric effect in presence of the external magnetic field. The situation is in this case  much more complex in comparison with a zero-magnetic-field case, because the overal number of ground states increases significantly.  Besides  the six previously  described ground states (given in Tab.~\ref{tab1}) we have additionally detected  eight different ground states, which can be defined  by combination of the  eigenvectors  given by Eq.~(\ref{eq5})  and the other two eigenvectors
\begin{align}
|\textrm{II}_2\rangle_k^{\delta}&=|1\rangle_{\sigma_{k,l_1}}\!\otimes\!\left(|\uparrow,\uparrow\rangle_k\right)\!\otimes\!|1\rangle_{\sigma_{k,l_2}}\;,
\nonumber\\
|\textrm{II}_3\rangle_k^{\delta}&=|1\rangle_{\sigma_{k,l_1}}\!\otimes\!\left[a_3^{\delta}(|\uparrow,\downarrow\rangle_k\!-\!|\downarrow,\uparrow\rangle_k)\!+\!b_3^{\delta}|\uparrow\downarrow,0\rangle_k\right.\nonumber\\
&+\left.c_3^{\delta}|0,\uparrow\downarrow\rangle_k\right]\!\otimes\!|1\rangle_{\sigma_{k,l_2}}\;,
\nonumber\\
a_3^{\delta}&=\frac{t}{2\sqrt{V_{\delta}^2\!+\!t^2}}\;,
\hspace*{0.7cm}
b_3^{\delta}=\frac{\sqrt{V_{\delta}^2\!+\!t^2}\!+\!V_{\delta}}{2\sqrt{V_{\delta}^2+t^2}}\;,\hspace*{0.7cm}
c_3^{\delta}=\frac{\sqrt{V_{\delta}^2\!+\!t^2}\!-\!V_{\delta}}{2\sqrt{V_{\delta}^2+t^2}}\;.
\label{eq6}
\end{align}
A complete list of eight novel  ground states is reported in Tab.~\ref{tab2}, 
\begin{table}[b!]
\begin{center}
\begin{tabular}{l|c|c|l|c}
\hline
\rowcolor[gray]{0.7}
$\rho$&{Phase} &${\varepsilon}_k$&$|\psi\rangle_k$&$m_{tot}$\\
\hline\hline
1 &G & $-h-2\mu-2\omega_x$&$|{\rm II}_3\rangle_k^x|{\rm 0}\rangle_k^y$&2/3\\
\hline
1.5&H &$-J-2h-3\mu-2\omega_x-\omega_y$& $|{\rm II}_3\rangle_k^x|{\rm I}\rangle_k^y$&2/3\\
1.5 &I&$-3J-4h-3\mu-\omega_x$& $|{\rm I}\rangle_k^x\;\;\,|{\rm II}_2\rangle_k^y$&1\\
\hline
2&J &$-2J-5h/2-4\mu-\Omega_x$ & $|{\rm II}\rangle_k^x\;\,|{\rm II}_2\rangle_k^y$&1/2\\
2&K &$-h/2-4\mu-2\omega_x-\Omega_y$& $|{\rm II}_3\rangle_k^x|{\rm II}\rangle_k^y$&1/6\\
2&L &$-2J-3h-4\mu-2\omega_x$& $|{\rm II}_3\rangle_k^x|{\rm II}_2\rangle_k^y$&2/3\\
2&M &$-h-4\mu-2\omega_x-2\omega_y$& $|{\rm II}_3\rangle_k^x|{\rm II}_3\rangle_k^y$&1/3\\
2&N &$-4J-5h-4\mu$& $|{\rm II}_2\rangle_k^x|{\rm II}_2\rangle_k^y$&1\\
\hline
\end{tabular}
\caption{\normalsize A list of additional eight  ground states of a coupled spin-electron  model on a doubly decorated square lattice emergent due to presence of the external magnetic field. Here ${\Omega_{x(y)}}\!=\!\sqrt{2(J^2\!+\!\omega_{x(y)}^2)\!+\!2\sqrt{(J^2\!+\!\omega_{x(y)}^2)^2\!-\!4J^2V_{x(y)}^2}}$ and ${\omega_{x(y)}}\!=\!\sqrt{V_{x(y)}^2\!+\!t^2}$\;.}
\label{tab2}
\end{center}
\end{table}
where individual ground states are supplemented with the definition of their notation, ground-state energy (${\varepsilon}_k$), the explicit form of the relevant eigenvector ($|\psi\rangle_k$) and the total magnetization ($m_{tot}$).
It can be shown that more than one possible ground state may emerge for any out of three considered electron concentrations $\rho\!=\!1.0$, 1.5 and 2.0. Therefore, the existence of the rotating magnetoelectric effect  under presence of the external magnetic field could be expected.  
The ground-state phase diagrams in the angle versus a relative  strength of the electric field ($\varphi-V/J$) plane under the influence of the non-zero magnetic field $h/J$ are presented in Figs.~\ref{fig4}-\ref{fig6}.
In agreement with general expectations   the applied magnetic field  stabilizes the ferromagnetic structures instead of the antiferromagnetic one at a quarter  filling ($\rho\!=\!1$), whereas the relevant ground-state phase boundaries  depend monotonically on the angular variable $\varphi$ (see Fig.~\ref{fig4}). Consequently, the applied magnetic field preserves the rotating magnetoelectric effect at a quarter electron filling, however, it shifts its occurrence to the higher value of the electric field $V/J$.

The investigated system at the mean electron concentration $\rho\!=\!1.5$ can exhibit at most three zero-temperature phase transitions driven by the electric field $V/J$, which may result  in four different types of  ground-state phase diagrams shown in Fig.~\ref{fig5}.  It was found that  the magnetic field dominantly influences the spin subsystem, which becomes ferromagnetic, although the changes in the electron subsystem strongly depend on a magnitude of the applied electric field. Similarly as in the previous case ($\rho\!=\!1$),  the non-negligible effect of a polar angle $\varphi$ determining a spatial orientation of the external electric field on a stability of the relevant ground states is also evident for the particular  case  $\rho\!=\!1.5$.  Moreover, the obtained  phase diagrams at  non-zero magnetic field $h/J\!\neq\!0$ and $\varphi\!<\!\pi/4$ demonstrate that the unconventional phenomenon of swapping  magnetic orders on the horizontal and vertical direction (E$\rightleftarrows$E$^*$) is dramatically reduced upon the applied magnetic field. However a complete suppression of this phenomenon is detected just above a  relatively large magnetic field $h/J\approx 0.9$.

In the most  interesting half-filled case $\rho\!=\!2$ the rotating magnetoelectric effect is substantially influenced by a  competition of both applied fields, because the non-zero magnetic field  favors six different  ground states instead of a single ground state F detected in zero-magnetic-field case.   Under this condition, there are three possible types of  ground-state phase diagrams as illustrated Fig.~\ref{fig6}. The most diverse ground-state  phase diagram can be found at very low magnetic fields, where the competition between the isotropic and anisotropic tendency of the magnetic and electric fields leads to the existence of six  different  phases. Of course, the phase boundaries between the individual ground states are substantially  affected by the  hopping term $t/J$, which generally stabilizes  spatially anisotropic structures instead of the isotropic ones. With no exception, all phase boundaries  exhibit a monotonic angular dependence demonstrating  presence of the rotating magnetoelectric effect. The higher values of the magnetic field $h/J$ as well as the   hopping amplitude $t/J$ reduce the overall structure of the ground-state phase diagram due to suppression of some phases. While the competing effect of moderate magnetic field (e.g. $h/J\!=\!0.2$) and the applied electric field still preserves  the electron segregation over the horizontal and vertical bonds, the sufficiently strong magnetic field (e.g. $h/J\!=\!1.0$) reorients spins into its direction and the homogeneous distribution of electrons becomes energetically favourable. In all identified phase boundaries the influence of the polar angle $\varphi$ specifying a spatial orientation of the external electric field is evident, and thus  existence of the rotating magnetoelectric effect is observed.

\section{Conclusion}
\label{conclusion}
The ground-state properties of a coupled  spin-electron model on a doubly decorated square lattice have been investigated under the influence of the external electric field acting either separately or in a mutual competition with an external magnetic field. The main goal was to study presence of the intriguing rotating magnetoelectric effect achieved upon the sample rotation in a homogeneous external electric field. It was found that the dissimilar impact of the electric field on the electron subsystems in the horizontal and vertical directions produces  novel inhomogeneous magnetic ground states, which become dominant upon  increasing a relative strength of the electrostatic potential  and/or decreasing the polar angle determining the spatial orientation of the external electric field. Due to this spatial anisotropy, the investigated spin-electron system in absence of an external magnetic field can exhibit two  different  ground states for the electron concentrations $\rho\!=\!1$ and $\rho\!=\!1.5$. 
 A monotonic (non-constant) character of a discontinuous phase transition  between  two respective   ground states driven  by a spatial orientation of the electric field directly confirms presence of a rotating magnetoelectric effect in a coupled spin-electron system on a doubly decorated square lattice.  
It turns out that the rotating magnetoelectric effect   also persists for both aforementioned electron concentrations ($\rho\!=\!1$ and $\rho\!=\!1.5$) in presence of the external magnetic field, which in addition generates novel spatially anisotropic  ground states. It was shown that the phase boundaries among all novel phases  very sensitively depend on  the polar angle $\varphi$ specifying a spatial orientation of the external electric field, what equally demonstrates presence of the rotating magnetoelectric effect. In addition, it was found that presence of an arbitrarily small but non-zero magnetic field  at a half filling $\rho\!=\!2$ generates at most five novel  ground states depending on a relative strength  of the magnetic field $h/J$ and the hopping therm $t/J$. A monotonic dependence of detected phase boundaries between each two coexisting phases on  the polar angle $\varphi$ confirm  existence of the rotating magnetoelectric effect also in the physically most interesting half-filled case $\rho\!=\!2$. It should be emphasized that this effect is a direct consequence of a competition between both applied external fields (magnetic as well as electric) and cannot exist in  absence of one of them. Besides, the variation of the polar angle $\varphi\!<\!\pi/4$ can produce another interesting behavior, exclusively observed at the fractional electron concentration $\rho\!=\!1.5$ and a relative weak  hopping therm $t/J$, where the magnetic orders of the spin-electron system on a doubly decorated square lattice can be easily swapped on the horizontal and vertical direction (E$\leftrightarrows$E$^*$).  Although the strengthening of the magnetic field  significantly reduces this phenomenon, it is necessary to apply sufficiently strong magnetic fields  $h/J\approx 0.9$ to fully suppress this swapping. 
Without any doubts, the aforementioned  diversity makes such spin-electron systems very attractive for a possible application in spintronics, sensorics or storage devices.

\vspace{0.5cm}
This work was supported by the Slovak Research and Development Agency (APVV) under Grant No. APVV-16-0186. The financial support provided by the VEGA under Grant No. 1/0105/20 is also gratefully acknowledged. 
\\


\begin{thebibliography}{00}
\bibitem{Fiebig} M. Fiebig, J. Phys. D {\bf 38}, R123 (2005).
\bibitem{Nikitin} S.~A. Nikitin, K.~P. Skokov, Y.~S. Koshkidko, Y.~G. Pastushenkov, and T.~I. Ivanova, Phys. Rev. Lett. {\bf 105} (2010) 137205.
\bibitem{Tokura} Y. Tokura, S. Seki, and N. Nagaosa, Rep. Prog. Phys. {\bf 77} (2014) 076501.
\bibitem{Cao} M.~S. Cao, X.~X. Wang, M. Zhang, J.~Ch. Shu, W.~Q. Cao, H.~J. Yang, X.~Y. Fang, and J. Yuan, Adv. Funct. Mater. {\bf 29} (2019) 1807398.
\bibitem{Prinz}G.~A. Prinz, Science {\bf 282} (1998) 1660.
\bibitem{Wolf} S.~A. Wolf, D.~D. Awschalom, R.~A. Buhrman, J.~M. Daughton, S. von Moln\'ar, M.~L. Roukes, A.~Y. Chtchelkanova, and D.~M. Treger, Science {\bf 294} (2001) 1488. 
\bibitem{Son} J.~Y. Son, J.-H. Lee, S. Song, Y.-H. Shin, and H.~M. Jang, ACS Nano {\bf 7} (2013) 5522.
\bibitem{Scott} J.~F. Scott, Nat. Mater. {\bf 6} (2007) 256.
\bibitem{Hur} N. Hur, S. Park, P. A. Sharma, J. S. Ahn, S. Guha, and S.-W. Cheong, Nature {\bf 429} (2004) 392.
\bibitem{Wu} S.~M. Wu, S.~A. Cybart, P. Yu, M.~D. Rossell, J.~X. Zhang, R. Ramesh, and R.~C. Dynes, Nat. Mater. {\bf 9} (2010) 756.
\bibitem{Vopson} M. M. Vopson, Crit. Rev. Solid State {\bf 40} (2015) 223.
\bibitem{Sreenivasulu} G. Sreenivasulu, P. Qu, V. Petrov, H. Qu, and G. Srinivasan, Sensors {\bf 16} (2016) 262.
\bibitem{Cenci2019}  H. \v Cen\v carikov\'a and J. Stre\v cka, Phys. Lett. A {\bf 383} (2019) 125957.
\bibitem{Zhang} H. Zhang, Y.~W. Li, E.~K. Liu, Y.~J. Ke, J.~L. Jin, Y. Long, and B.~G. Shen, Scie. Rep. {\bf 5} (2015) 11929.
\bibitem{Caro} J. Caro Pantino and N.~A. de Oliveira, Intermetallics {\bf 64} (2015) 59.
\bibitem{Lorusso} G. Lorusso, O. Roubeau, M. Evangelisti, Angew. Chem. {\bf 55} (2016) 3360.
\bibitem{Balli1} M. Balli, S. Jandl, P. Fournier, and F.~Z. Dimitrov,  Appl. Phys. Lett. {\bf 108} (2016) 102401.
\bibitem{Balli3} M. Balli, S. Jandl, P. Fournier, and  A. Kedous-Lebouc, Appl. Phys. Rev. {\bf 4} (2017) 021305.
\bibitem{Orendac} M. Orenda\v{c}, S.  Gab\'ani, E. Ga\v{z}o, G. Prist\'a\v{s}, N. Shitsevalova, K. Siemensmeyer, and K. Flachbart, Scie. Rep. {\bf 8} (2018) 10933.
\bibitem{Moon} J.~Y. Moon, M.~K. Kim, D.~G. Oh, J.~H. Kim, H.~J. Shin,Y.~J. Choi, and N. Lee, Phys. Rev. B {\bf 98} (2018) 174424.
\bibitem{Cenci1} H. \v Cen\v carikov\'a, J. Stre\v cka, and A. Gendiar, J. Magn. Magn. Mater. {\bf 452} (2018) 512.
\bibitem{Cenci2} H. \v Cen\v carikov\'a, J. Stre\v cka, and M. L. Lyra, J. Magn. Magn. Mater. {\bf 401} (2016) 1106.
\bibitem{Cenci4} H. \v Cen\v carikov\'a and J. Stre\v cka, Phys. Rev. E {\bf 98} (2018) 062129.
\bibitem{Cenci0} H. \v Cen\v carikov\'a, J. Stre\v cka, A. Gendiar, and N. Toma\v sovi\v cov\'a, Physica B {\bf 536} (2018) 432.
\bibitem{Cenci2015} J. Stre\v{c}ka, H. \v{C}en\v{c}arikov\'a, M. L. Lyra, Phys. Lett. A {\bf 379} (2015) 2915.

\end{thebibliography}
\end{document}